\documentclass[aps,prd,floats,floatfix,twoside,preprintnumbers,superscriptaddress,nofootinbib]{revtex4}

\usepackage{graphicx} 
\usepackage{dcolumn}
\usepackage{amssymb}
\usepackage{amsmath}
\usepackage[hidelinks]{hyperref}
\usepackage{color}
\usepackage[utf8]{inputenc}

\newcommand{\xp}{\boldsymbol{x}_\perp} 

\newcommand{\kp}{\boldsymbol{k}_\perp}

\begin{document}

\preprint{YITP-18-118}

\date{\today}

\title{Single spin asymmetry in forward $pA$ collisions: Phenomenology at RHIC}
\author{Sanjin Beni\' c\footnote{On leave of absence from: Department of Physics, Faculty of Science, University of Zagreb, Bijeni\v cka c. 32, 10000 Zagreb, Croatia}}
\affiliation{Yukawa Institute for Theoretical Physics,
Kyoto University, Kyoto 606-8502, Japan}

\author{Yoshitaka Hatta}
\affiliation{Physics Department, Brookhaven National Laboratory, Upton, New York 11973, USA}
\affiliation{Yukawa Institute for Theoretical Physics,
Kyoto University, Kyoto 606-8502, Japan}

\begin{abstract} 
We confront the theoretical result of single spin asymmetry (SSA) $A_N$ in forward $pA$ collisions $p^\uparrow A \to hX$ including the gluon saturation effect with the recent preliminary experimental data from the PHENIX and STAR collaborations at RHIC. While we find  overall reasonable agreement with the STAR data, our results indicate that the strong nuclear suppression of the asymmetry $A_N\sim A^{-1/3}$ observed by the PHENIX collaboration cannot be explained within the present understanding of this problem. 
\end{abstract}


\maketitle 

\section{Introduction}

Transverse single spin asymmetries (SSAs), as measured in collisions of an unpolarized probe with a transversely polarized proton, are traditionally a venue to understand the spin structure of the proton \cite{DAlesio:2007bjf,Boer:2011fh,Perdekamp:2015vwa}. For inclusive hadron productions at high-$P_{h\perp}$, SSA is computed from perturbative QCD where it becomes a probe of collinear twist-3 distributions.
Recent measurements at RHIC considered collisions of polarized protons on nuclear targets and so a completely new interplay between spin physics and the physics of gluon saturation becomes a reality 
 \cite{Boer:2006rj,Kang:2011ni,Kang:2012vm,Kovchegov:2012ga,Altinoluk:2014oxa,Schafer:2014zea,Zhou:2015ima,Boer:2015pni}. This is especially so, as gluon saturation is important in the forward region of the produced hadron where SSA is the largest.
Both the STAR \cite{Dilks:2016ufy} and the PHENIX collaborations \cite{Aidala:2019ctp}   reported on preliminary results  on SSA in $p^\uparrow A \to h X$ in addition to $p^\uparrow p \to h X$. The PHENIX collaboration found a striking nuclear suppression $A_N\propto A^{-1/3}$ with the mass number $A$ in their preliminary data sets. On the other hand, the STAR collaboration did not find any significant nuclear effect in their data. The two data sets are not necessarily in contradiction to each other, as they are collected for different kinematics. However, the difference in kinematics is actually not very large, and both data are sensitive to the small-$x$ region of the nucleus target. Therefore, it remains a challenge for theorists to explain both data consistently in a single framework. 

At first sight,   the suppression $A_N\propto A^{-1/3}$ seems consistent with the prediction  of $k_\perp$-factorization approaches \cite{Boer:2006rj,Kang:2011ni}  which include the gluon saturation effects \cite{Iancu:2003xm,Gelis:2010nm} in the target nucleus. However, the $k_\perp$-factorization does not apply to this process, and a  more proper treatment based on the collinear or `hybrid' \cite{Schafer:2014zea} factorization has identified two contributions with different scaling behaviors $A_N\sim {\cal O}(A^0) +{\cal O}(A^{-1/3})$  \cite{Hatta:2016wjz,Hatta:2016khv,Zhou:2017sdx}. The recent fits of the $p^\uparrow p$ data \cite{Kanazawa:2014dca,Gamberg:2017gle} indicate that the ${\cal O}(A^{0})$ terms are dominant, so the PHENIX result is actually surprising.  Furthermore, the suppression is observed at  relatively high values of the hadron $P_{h\perp}$ where one does not expect to see strong nuclear effects at RHIC energies. In view of this, it is premature to link the PHENIX finding with the gluon saturation effect. 


In this paper we quantitatively address this problem by numerically computing the SSA in $p^\uparrow p \to h X$ and $p^\uparrow A \to h X$ using the  formulas derived in \cite{Hatta:2016wjz,Hatta:2016khv}. We then compare our results with the preliminary STAR and  PHENIX data. Following \cite{Kanazawa:2014dca,Gamberg:2017gle}, we assume that the twist-3 fragmentation contribution is the main cause of SSA in this channel. As for the nucleus, we use the solution 
of the running coupling Balitsky-Kovchegov (rcBK) equation \cite{Balitsky:1995ub,Kovchegov:1999yj}. The main result, presented in Sec.~\ref{sec:res} shows an overall satisfactory agreement with the preliminary STAR data. On the other hand, we were not able to confirm the nuclear suppression as seen in the PHENIX preliminary results. To the contrary, our results show no nuclear dependence for the PHENIX kinematics, even though we include the saturation effect of the target. We  investigate the reason of this failure and discuss what extra contributions are needed to fix this problem. 

\section{Cross section formulas}
\label{sec:form}

Our starting point are the formulas for the spin independent $pA \to hX$ cross section and the fragmentation contribution to the spin dependent $p^\uparrow A \to hX$ cross section within the CGC framework. 
The spin independent cross section \cite{Dumitru:2005gt} is given as
\begin{equation}
\frac{d\sigma}{d^2 P_{h\perp} dy_h} = \sum_a \int_{z_{\rm min}}^1 \frac{dz}{z^2}x_q f_a(x_q,P_{h\perp}^2)F(x_g,P_{h\perp}/z)D_{h/a}(z,P_{h\perp}^2)\,,
\label{eq:xs}
\end{equation}
where $P_{h\perp}$ and $y_h$ are the hadron transverse momenta and rapidity, respectively, with $x_{q,g} = \frac{P_{h\perp}}{z\sqrt{s}}e^{\pm y_h}$ and $z_{\rm min} = \frac{P_{h\perp}}{\sqrt{s}}e^{y_h}$ and we sum over quark flavors as $\sum_a$.
The function $F(x,k_\perp)$ is the gluon dipole distribution defined as
\begin{equation}
F(x,k_\perp) \equiv \pi R_A^2\int \frac{d^2 \xp}{(2\pi)^2} e^{-i\kp \cdot \xp} F_{Y}(x_\perp) \equiv \pi R_A^2\int \frac{d^2 \xp}{(2\pi)^2} e^{-i\kp \cdot \xp}\frac{1}{N_c}\left\langle \mathrm{Tr}\left[U(\xp) U^\dag(0)\right]\right\rangle_{Y}\,,
\label{eq:dip}
\end{equation}
where $Y = \log(1/x)$, $R_A$ is the nuclear radius and $U(\xp)$ is the fundamental Wilson line with $\langle \dots\rangle$ in the third line denoting the color average. In Eq.~\eqref{eq:xs} $f_a(x,Q^2)$ is the unpolarized parton distribution function and $D_{h/a}(z,Q^2)$ is the unpolarized hadron fragmentation function evaluated at the scale $Q^2 = P_{h\perp}^2$.

The spin dependent cross section comes from the quark-gluon-quark contribution, the twist-3 fragmentation contribution and the triple gluon contribution. In this work we consider only the fragmentation contribution as it is the dominant source of SSA in this channel \cite{Kanazawa:2014dca,Gamberg:2017gle}. We start from the main formula (see Eq.~(46) in \cite{Hatta:2016khv})
\begin{equation}
\begin{split}
\frac{d\Delta\sigma(S_\perp)}{d^2 P_{h\perp} dy_h} &= \frac{M}{2} S_{\perp i}\epsilon^{ij}\sum_a\int_{z_{\rm min}}^1\frac{dz}{z^2} x_q h^a_1(x_q,P_{h\perp}^2)\Bigg\{-\mathrm{Im} \tilde{e}^{h/a}(z,P_{h\perp}^2) \frac{d}{d (P_h^j/z)}F(x_g,P_{h\perp}/z)\\
& + 4 \frac{P_{hj}}{P_{h\perp}^2}\int_{z}^\infty \frac{d z_1}{z_1^2}\frac{z}{\frac{1}{z} - \frac{1}{z_1}}\frac{\mathrm{Im} \hat{E}^{h/a}_F(z_1,z,P_{h\perp}^2)}{N_c^2 - 1}\left[\frac{2\pi N_c^2}{\pi R_A^2} \int_0^{P_{h\perp}/z_1}l_\perp d l_\perp F(x_g,l_\perp) + \frac{1}{z_1}\frac{1}{\frac{1}{z} - \frac{1}{z_1}}\right]F(x_g,P_{h\perp}/z)\Bigg\}\,,
\end{split}
\label{eq:xsspin2}
\end{equation}
where $M$ is the proton mass and $S_\perp^i$ is the proton spin. Here $h_1^a(x,Q^2)$ is the quark transversity distribution, while $\mathrm{Im}\tilde{e}^{h/a}(z,Q^2)$ and $\mathrm{Im} \hat{E}^{h/a}_F(z',z,Q^2)$ are the hadron twist-3 fragmentation functions.

In the next step we approximate $\int_0^{P_{h\perp}/z_1} dl_\perp \simeq \int_0^{P_{h\perp}/z} d l_\perp$ which is reasonable considering that $z_1 >z$ while $F(x,k_\perp)$ is a monotonically dropping function of $k_\perp$. With this approximation it is possible to make use of the following relations for the twist-3 fragmentation functions
\begin{equation}
\begin{split}
&\hat{e}^{h/a}_{\bar{1}}(z,Q^2) = z\mathrm{Im} \tilde{e}^{h/a} (z,Q^2) + z\int_z^\infty \frac{dz'}{z'^2}P\left(\frac{1}{1/z' - 1/z}\right)\mathrm{Im} \hat{E}^{h/a}_F(z',z,Q^2)\,,\\
& \frac{\hat{e}^{h/a}_{\bar{1}}(z,Q^2)}{z} = \frac{1}{2}\frac{d}{d(1/z)}\left(\frac{\mathrm{Im} \tilde{e}^{h/a} (z,Q^2)}{z}\right) + \frac{1}{z}\int_z^\infty \frac{dz'}{z'^2}P\left(\frac{1}{(1/z - 1/z')^2}\right)\mathrm{Im} \hat{E}^{h/a}_F(z',z,Q^2)\,,
\end{split}
\label{eq:ft3}
\end{equation}
to eliminate the terms containing the $z'$ integral over $\mathrm{Im} \hat{E}^{h/a}_F(z',z,Q^2)$ in \eqref{eq:xsspin2}. Here $\hat{e}^{h/a}_{\bar{1}}(z,Q^2)$ is yet another twist-3 fragmentation function. The notation used in this work relates to the notation in Ref.~\cite{Kanazawa:2015ajw,Gamberg:2017gle} as
\begin{equation}
H^{h/a}(z,Q^2) = - \frac{M_N}{M_h} \hat{e}_{\bar{1}}^{h/a}(z,Q^2) \,, \quad 
H^{\perp (1),h/a}(z,Q^2) = \frac{M_N}{2 M_h} \mathrm{Im}\tilde{e}^{h/a}(z,Q^2) \,, \quad \mathrm{Im}\hat{H}_{FU}^{h/a}(z,z',Q^2) = \frac{M_N}{2 M_h}\mathrm{Im} \hat{E}_F^{h/a}(z',z,Q^2)\,,
\label{eq:Hs}
\end{equation}
with $H^{h/a}(z,Q^2)$, $H^{\perp(1),h/a}(z,Q^2)$ and $\hat{H}^{h/a}_{FU}(z,z',Q^2)$ named as intrinsic, kinematical, and the dynamical twist-3 fragmentation functions.
The equations \eqref{eq:ft3} are known as the QCD equation of motion relation \cite{Kang:2010zzb,Metz:2012ct} and the Lorentz invariance relation \cite{Kanazawa:2015ajw}, respectively. 
Using \eqref{eq:ft3}, Eq.~\eqref{eq:xsspin2} becomes
\begin{equation}
\begin{split}
\frac{d\Delta \sigma(S_\perp)}{d^2 P_{h\perp} dy_h} &= \frac{M}{2} S_{\perp i}\frac{P_{hj}}{P_{h\perp}}\epsilon^{ij}\sum_a\int_{z_{\rm min}}^1\frac{dz}{z^2} x_q h_1^a(x_q,P_{h\perp}^2)\Bigg\{\mathrm{Im} \tilde{e}^{h/a}(z,P_{h\perp}^2) \frac{d}{d (P_{h\perp}/z)}F(x_g,P_{h\perp}/z)\\
& + \frac{4}{P_{h\perp}}\frac{1}{N_c^2 - 1}\Bigg[\frac{2\pi N_c^2}{\pi R_A^2} \int_0^{P_{h\perp}/z}l_\perp d l_\perp F(x_g,l_\perp)\left(z\mathrm{Im}\tilde{e}^{h/a}(z,P_{h\perp}^2) - \hat{e}^a_{\bar{1}}(z,P_{h\perp}^2)\right) + 2 \hat{e}^{h/a}_{\bar{1}}(z,P_{h\perp}^2)\\
& - z\mathrm{Im} \tilde{e}^{h/a} (z,P_{h\perp}^2) - \frac{z}{2}\frac{d}{d(1/z)}\left(\frac{\mathrm{Im} \tilde{e}^{h/a} (z,P_{h\perp}^2)}{z}\right)\Bigg]F(x_g,P_{h\perp}/z)\Bigg\}\,.
\end{split}
\label{eq:xdep3}
\end{equation}

In the following we will numerically compute the SSA defined as
\begin{equation}
A_N = \frac{1}{2}\frac{d\Delta \sigma(\uparrow) - d\Delta \sigma(\downarrow)}{d\sigma}\,.
\end{equation}
where in the numerator (denominator) we have the spin dependent (independent) cross section defined by Eq.~\eqref{eq:xdep3} (Eq.~\eqref{eq:xs}). We adopt the convention by which $S_i P_{hj} \epsilon^{ij} /P_{h\perp} = \sin(\phi_h - \phi_S) = -1$, where $\phi_h$ ($\phi_S$) are azimuthal angles of the outgoing hadron (spin). When the incoming proton is pointing in the $+z$ direction, and with its spin pointing in the $y$ direction, $\Delta \sigma(\uparrow)$ ($\Delta \sigma(\downarrow) = -\Delta \sigma(\uparrow)$) is the cross section for the hadron emission in the $+x$ ($-x$) direction, or left (right) direction. This explains the ``left - right" convention which is also used by STAR and PHENIX.

The nuclear effects are contained in the dipole function $F(x,k_\perp)$ and especially the first term in the spin dependent cross section \eqref{eq:xdep3} depends on the derivative of the dipole. In the saturation regime ($k_\perp \lesssim Q_S$), where $Q_S$ is the saturation scale, we would typically get $dF/d k_\perp \sim k_\perp F/Q_S^2$. Since the spin independent cross section \eqref{eq:xs} goes as $\sim F(x,P_{h\perp}/z)$, we find, for this particular term,  $A_N\sim Q_S^{-2}$, leading to $A_N \sim A^{-1/3}$ for the nuclei. Although not immediately obvious, the second term of \eqref{eq:xdep3} also scales as $A^{-1/3}$ \cite{Hatta:2016khv}. From a quantitative perspective it is important that the saturation scale in the nuclei scales as $(Q_S^A)^2 = c A^{1/3} (Q_S^p)^2$ ($Q_S^p$ is the saturation scale in the proton) where an additional proportionality factor  $c<1$ \cite{Dusling:2009ni} (in the numerical calculations we will use $c = 0.5$) will inhibit the overall magnitude of the nuclear suppression.
On the other hand, when $k_\perp \gg Q_S$ we are in the perturbative regime where the dipole distribution has a characteristic dependence $F \sim Q_S^2/k_\perp^4$ and so $dF/d P_{h\perp} \sim Q_S^2/P_{h\perp}^5$. The same $Q_S$-dependence is found also for the second term in \eqref{eq:xdep3}: $F/P_{h\perp} \sim Q_S^2/P_{h\perp}^5$ and so in the perturbative limit the nuclear dependence drops out in the ratio.


\section{Calculation setup and numerical results}
\label{sec:res}

In this Section we first explain all the details of our calculation and then we numerically compute SSA
and compare with the available preliminary data from STAR and PHENIX.
We will often be using the Feynman-x variable: $x_F = 2 P_{h\perp}\sinh y_h/\sqrt{s}$.

For the dipole gluon distributions $F(x,k_\perp)$ \eqref{eq:dip}, we use the numerical solution of the running coupling Balitsky-Kovchegov (rcBK) equation \cite{Balitsky:1995ub,Kovchegov:1999yj} from \cite{Dusling:2009ni}. We take the McLerran-Venugopalan (MV) initial condition at $Y_0 = \log(1/x_0)$, where $x_0 = 0.01$ as
\begin{equation}
F_{Y_{p,A} = Y_0}(x_\perp) = \exp\left[-\frac{(x_\perp^2 (Q_{S,0}^{p,A})^2) ^\gamma}{4}\log\left(\frac{1}{x_\perp \Lambda} + e\right)\right]\,.
\label{eq:N0}
\end{equation}
Here $Q_{S,0}^{p,A}$ is initial saturation scale parameter for the proton and the nuclei, $\gamma$ is the anomalous dimension and $\Lambda$ the IR cutoff.
We use two different parameter sets for the initial condition. Labeling them as set MV and set MV$^\gamma$, the model parameters are: MV: $\gamma = 1$, $(Q^p_{S,0})^2 = 0.2$ GeV$^2$ and $\Lambda = 0.241$ GeV, MV$^\gamma$: $\gamma = 1.119$, $(Q_{S,0}^p)^2 = 0.168$ GeV$^2$ and $\Lambda = 0.241$ GeV. 
For the nuclei, we use the relation $(Q^A_{S,0})^2 = c A^{1/3}(Q^p_{S,0})^2$ with $c = 0.5$, as mentioned previously.

For the twist-2 distribution functions $f_a(x,Q^2)$, we use the central CTEQ10 set \cite{Lai:2010vv}. For the twist-2 pion and kaon fragmentation functions $D_{\pi/a}(z,Q^2)$ and $D_{K/a}(z,Q^2)$ we use the central DSSV set \cite{deFlorian:2014xna}.
The transversity distribution $h_1^a(x,Q^2)$ and the twist-3 pion fragmentation functions $\mathrm{Im} \tilde{e}^{\pi^+/a}(z,Q^2)$ are obtained by solving their respective evolution equations numerically with the initial condition determined in \cite{Kang:2015msa}. The twist-3 kaon fragmentation function is obtained from \cite{Anselmino:2015fty}. In both cases we employ the Wilczek-Wandzura approximation $\hat{e}^{h/a}_{\bar{1}}(z,Q^2) = z\mathrm{Im} \tilde{e}^{h/a} (z,Q^2)$ as in \cite{Kang:2015msa}.

\begin{figure}
  \begin{center}
  \includegraphics[scale = 0.18]{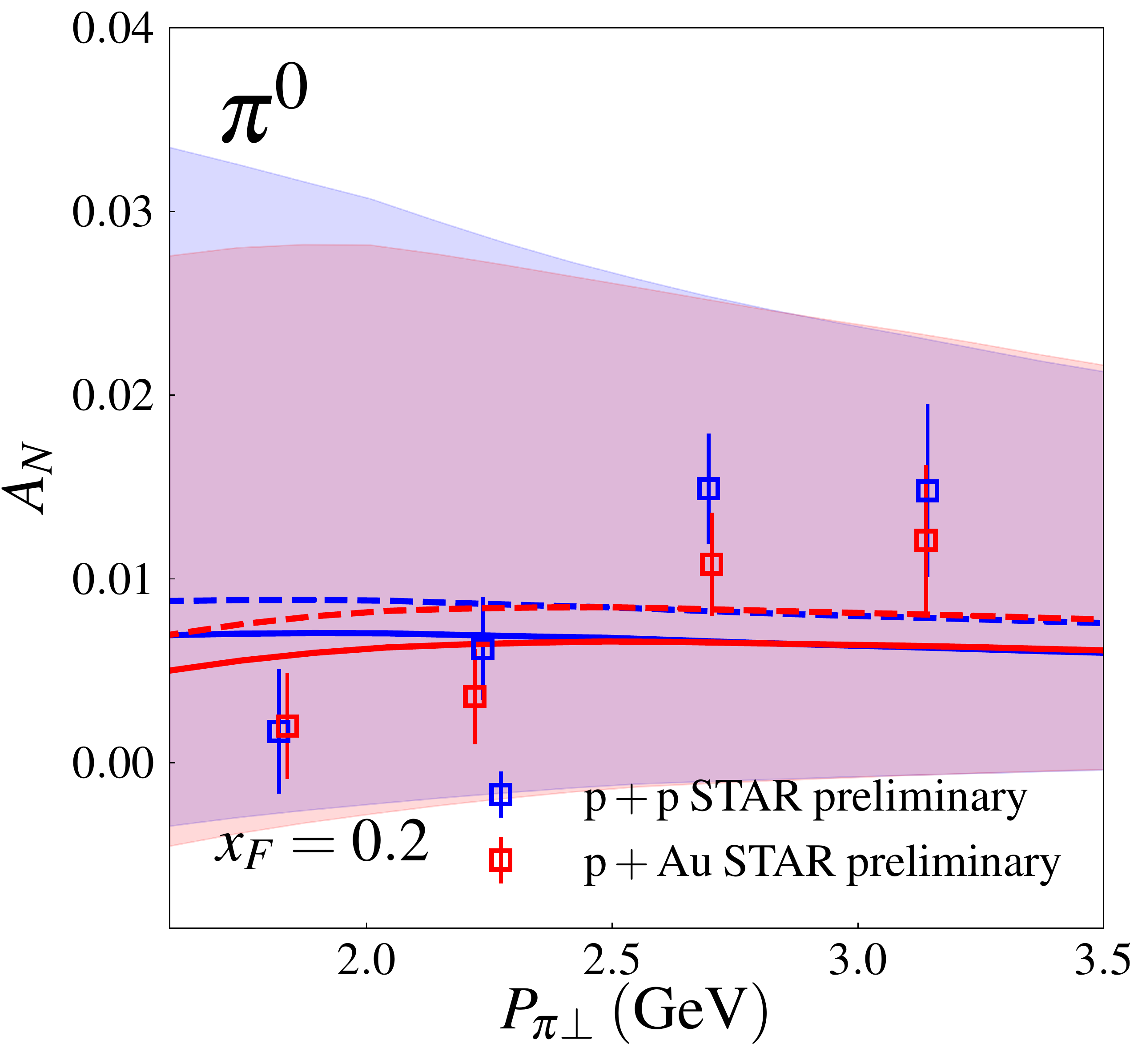}
    \includegraphics[scale = 0.18]{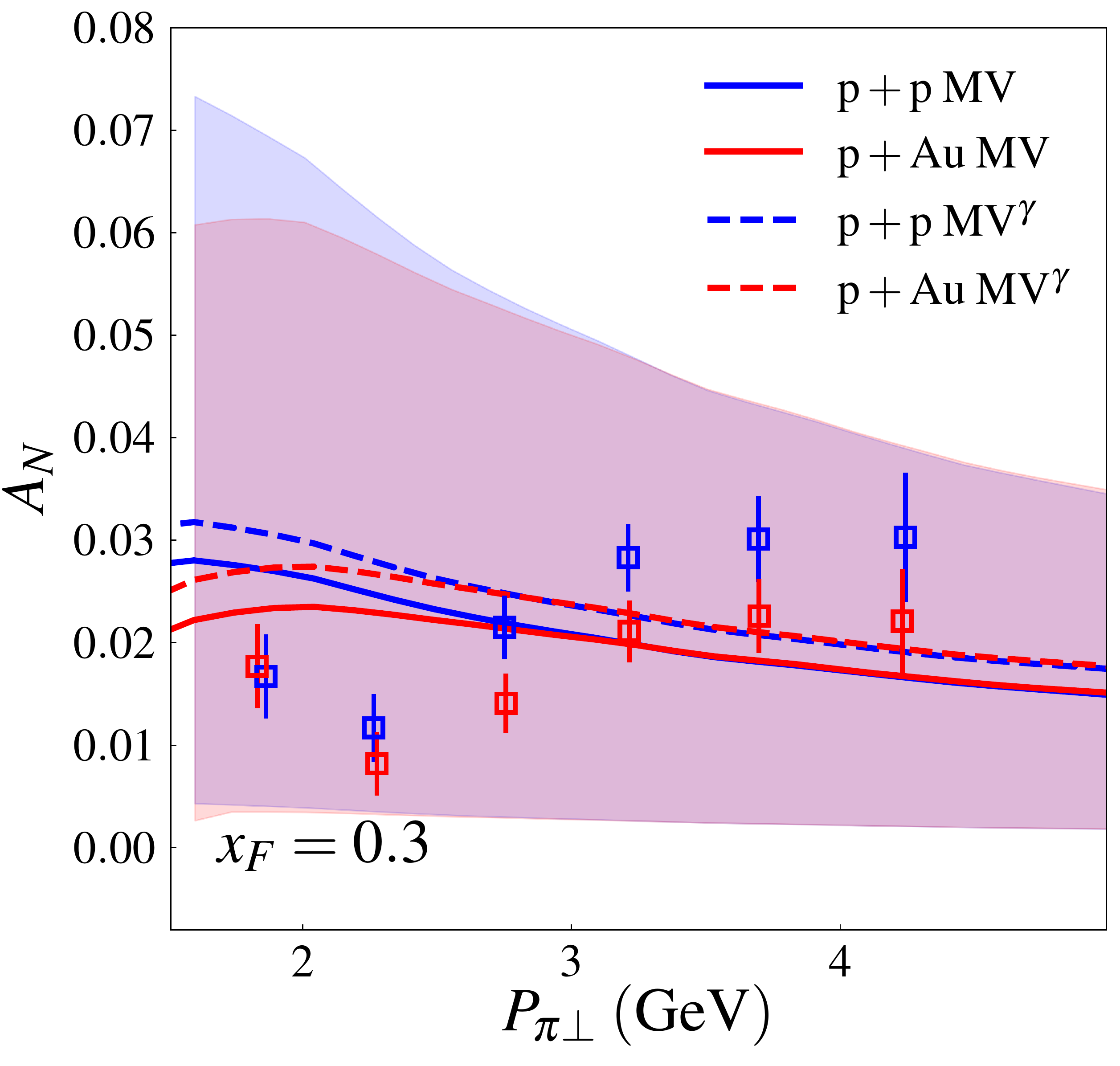}
      \includegraphics[scale = 0.18]{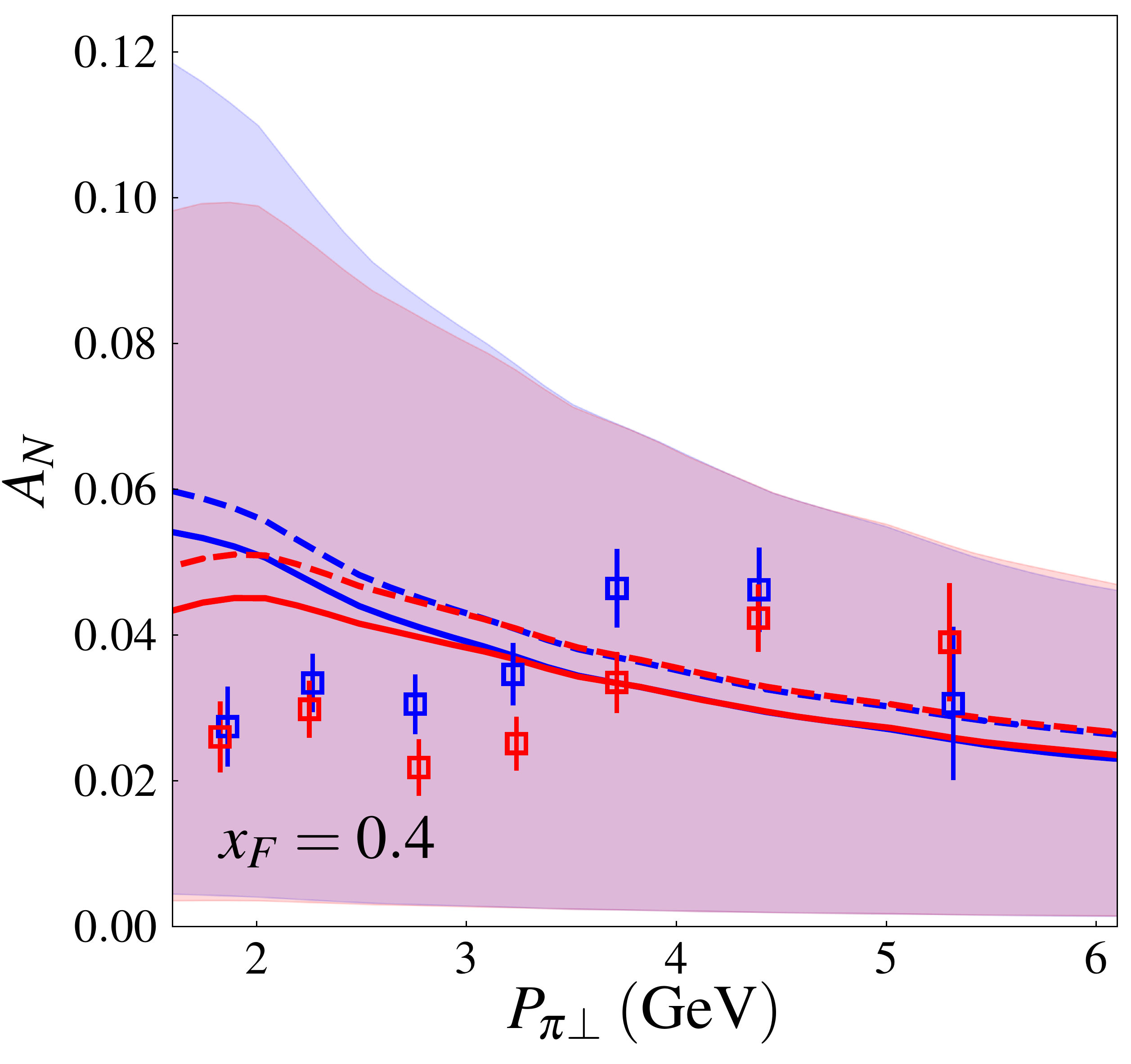}
        \includegraphics[scale = 0.18]{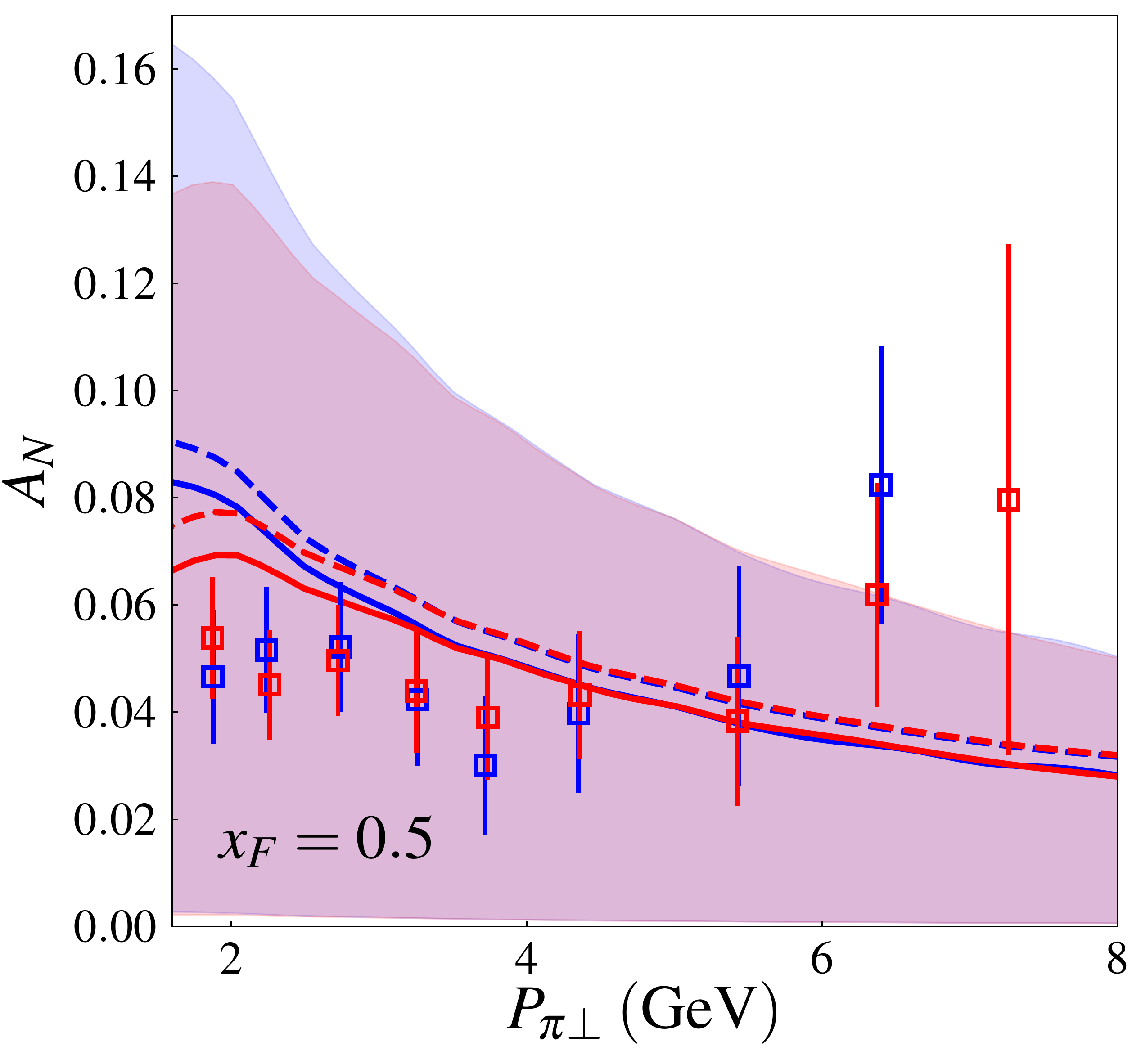}
    \includegraphics[scale = 0.18]{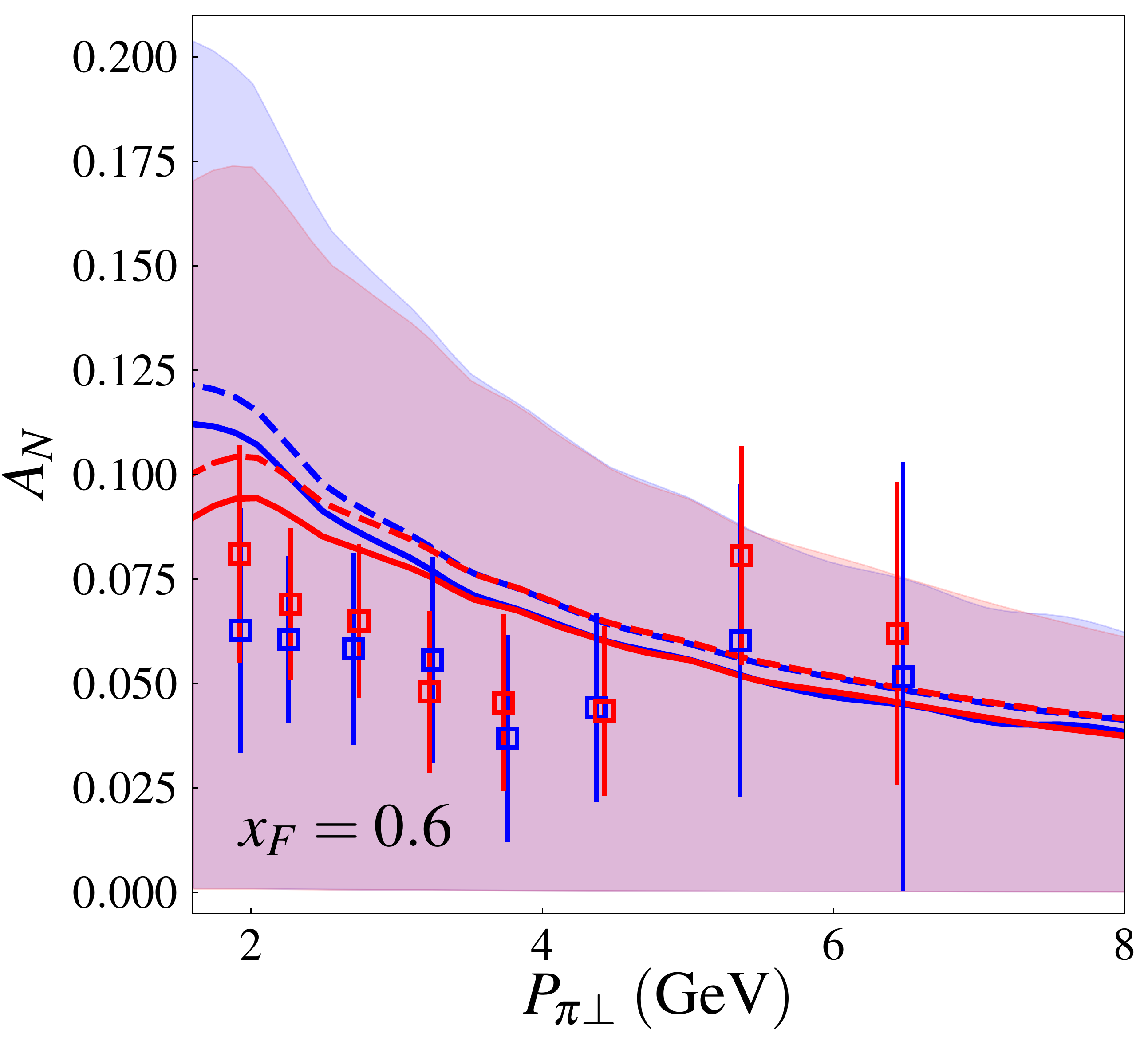}
      \includegraphics[scale = 0.18]{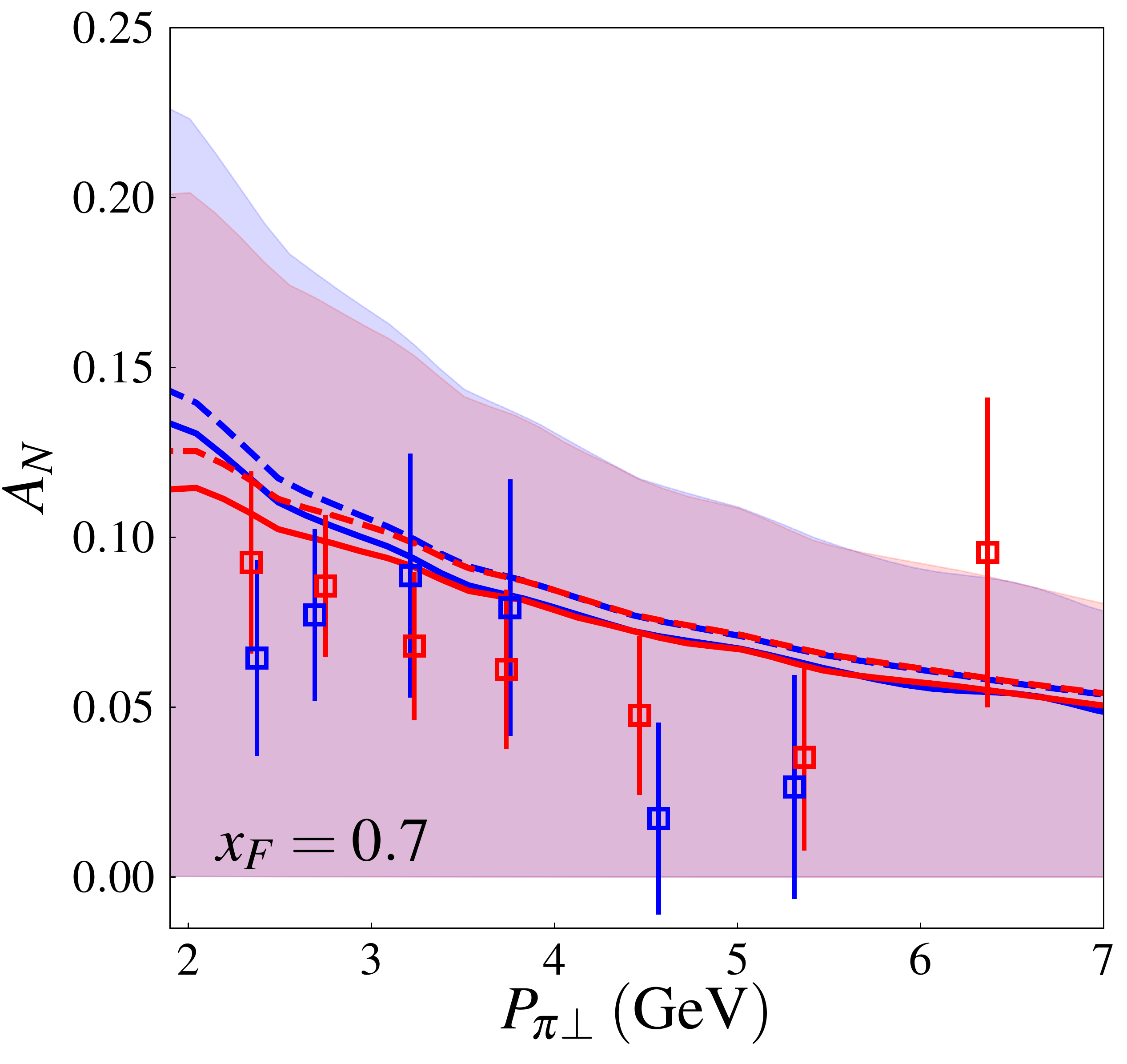}
  \end{center}
  \caption{$A_N$ for $p^\uparrow p\to \pi^0 X$ and $p^\uparrow Au\to \pi^0 X$ as a function of $P_{\pi\perp}$ at $\sqrt{s} = 200$ GeV versus preliminary STAR data \cite{Dilks:2016ufy}. The full (dashed) lines are a calculation using the MV (MV$^\gamma$) model where we used $(Q_{S,0}^{Au})^2 = 3(Q_{S,0}^p)^2$ as well as the central values for the transversity and the twist-3 fragmentation functions, while the shaded band reflects the uncertainty in extraction of both of these quantities according to \cite{Kang:2015msa}.}
  \label{fig:star}
\end{figure}

In Fig.~\ref{fig:star} we show the numerical results of our computation for $A_N$ in $p^\uparrow p \to \pi^0 X$ and $p^\uparrow Au \to \pi^0 X$ as a function $P_{\pi\perp}$ for several values of $x_F$ as compared to the preliminary STAR data \cite{Dilks:2016ufy}. We have used $(Q_{S,0}^{Au})^2 = 3 (Q_{S,0}^p)^2$. The full (dashed) lines correspond to the calculation with the MV (MV$^\gamma$) model. The shaded band comes from the uncertainty in the transversity and the twist-3 fragmentation function from the analysis in \cite{Kang:2015msa}. In this case, we have calculated $A_N$ with the MV model. As a consistency check, we have also computed $A_N$ in $p^\uparrow p$ using the collinear gluon PDF for the unpolarized proton (as was done in \cite{Gamberg:2017gle}). The result is in good agreement with the one from the rcBK solution shown in  Fig.~\ref{fig:star}.\footnote{Incidentally, we also confirmed that the forward approximation $\hat{s}\approx -\hat{u} \gg |\hat{t}|$  in the partonic subprocess used to derive the formula (\ref{eq:xsspin2}) is actually very good in the kinematics we consider.} 

While the central results, given by the full and dashed lines on Fig.~1, seem to compare well with the overall magnitude of the preliminary STAR data and within its experimental uncertainties, the shaded bands reflect a large theoretical uncertainty in the extraction of the transversity and the twist-3 fragmentation function.
Nonetheless, there is a valuable point to be made here regarding the nuclear dependence of $A_N$.
The nuclear suppression of $A_N$ for $p^\uparrow Au$ relative to $A_N$ in $p^\uparrow p$ that is visible in the low $P_{\pi \perp}$ region is not only the most prominent feature of our calculation, but also quite robust, being of a similar magnitude for the central results as well as for the shaded regions. On the other hand, there are no clear indications of nuclear effects in the STAR data, and in particular there is no nuclear suppression in the low $P_{\pi\perp}$ region as predicted here by the hybrid factorization framework. The slight hints of nuclear suppression in the higher $P_{\pi\perp}$ bins, visible on Fig.~\ref{fig:star} for $x_F = 0.3$ and $x_F = 0.4$ (within errors), and even enhancement for $x_F = 0.6$ (central values) cannot be reproduced with the present framework either.

\begin{figure}
  \begin{center}
  \includegraphics[scale = 0.3]{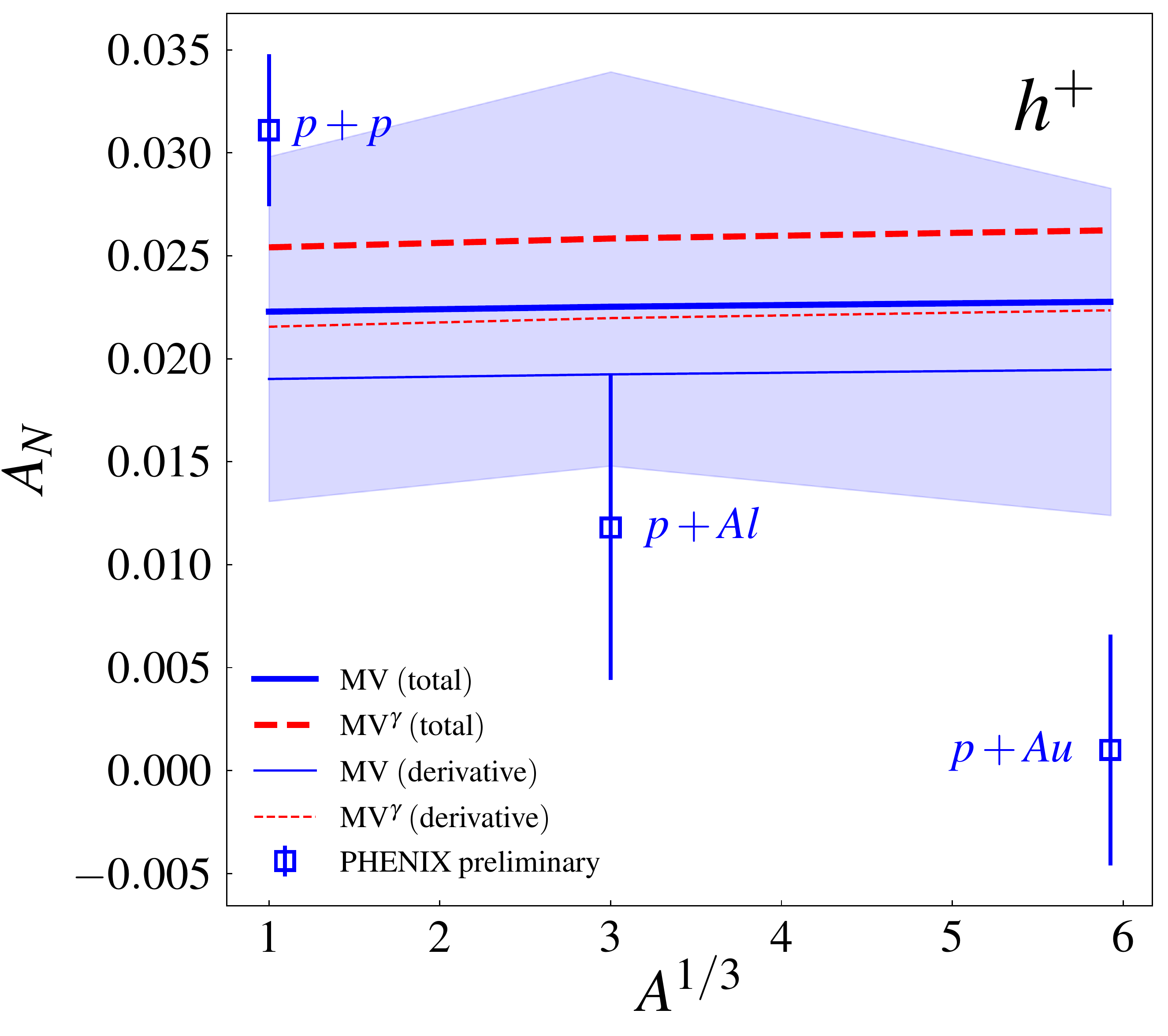}
  \end{center}
  \caption{$A_N$ as a function of $A^{1/3}$ at $\sqrt{s} = 200$ GeV versus preliminary data from PHENIX. The thick full (thick dashed) curves lines represent the MV (MV$^\gamma$) model calculation. In the case of the thin curves, we have taken into account only the derivative term in the polarized cross section (first term in Eq.~\eqref{eq:xdep3}) as a contribution to $A_N$. The shaded blue band takes into account the uncertainty of the transversity and the twist-3 fragmentation function according to \cite{Kang:2015msa}.}
  \label{fig:phenix}
\end{figure}

In the preliminary PHENIX data set \cite{Aidala:2019ctp}, covering a kinematics range $x_F \leq 0.12$, $A_N$ is measured in $p^\uparrow p \to h^+ X$, $p^\uparrow Al \to h^+ X$ and $p^\uparrow Au \to h^+ X$ where $h^+$ is a mixture of outgoing $\pi^+$ and $K^+$. The nuclear dependence of $A_N$ in the PHENIX data is most noticeable for the largest measured $x_F = 0.12$ where also an average $P_{h\perp} = 2.9$ GeV is measured. At these kinematics, $y_h = 2.13$ leading to $x_q > 0.12$, $x_g > 0.0017$ and so it is reasonable to apply the forward CGC formulas to compute $A_N$. 
In Fig.~\ref{fig:phenix} we show a comparison of our results for the nuclear dependence of $A_N$ in $p^\uparrow A \to h^+ X$\footnote{Even though our numerical calculation includes the contributions of $\pi^+$ and $K^+$, however, quantitatively the $K^+$ contribution is not more than about $10\%$ of the full result for $A_N$.} to the preliminary PHENIX data, for the kinematics point $x_F = 0.12$, $P_{h\perp} = 2.9$ GeV. In the PHENIX result $A_N$ clearly drops with the increase in the atomic number $A$, and this is consistent with the  behavior $A_N\sim A^{-1/3}$. However, our current numerical results show virtually no $A$-dependence. The  reason is clear: $P_{h\perp}=2.9$ GeV is too hard to be sensitive to the saturation scale which is $Q_S^{Au} \sim 0.9$ GeV for the PHENIX kinematics in the model used here.

\begin{figure}
  \begin{center}
  \includegraphics[scale = 0.3]{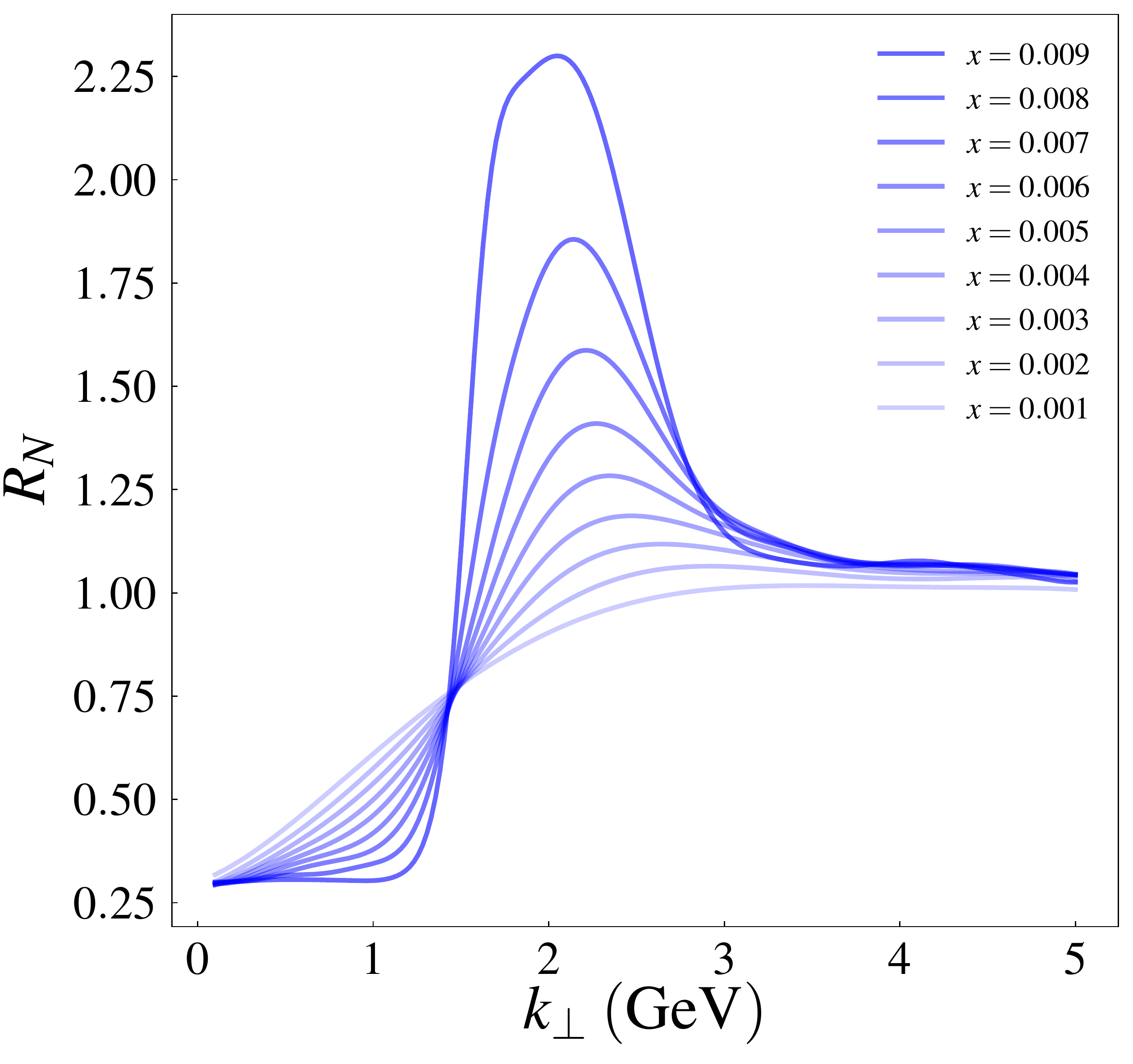}
  \end{center}
  \caption{The quantity $R_N$ (see Eq.~\eqref{eq:double}) for the definition) as a function of $k_\perp$ for different values of $x$.}
  \label{fig:double}
\end{figure}

To elaborate on this point, let us make an extreme assumption that only the first term of \eqref{eq:xdep3} is important. For $P_{hT}<Q_S^A$, this term is expected to give the scaling $A_N\sim A^{-1/3}$ and this is demonstrated in Fig.~\ref{fig:double} where we plot the double ratio 
\begin{equation}
R_N \equiv \frac{\left(dF/dk_\perp/F\right)_A}{\left(dF/dk_\perp/F\right)_p}\,.
\label{eq:double}
\end{equation}
as a function of $k_\perp$ for several values of $x$ using $(Q_{S,0}^{Au})^2 = 3 (Q_{S,0}^p)^2$. Close to the initial condition the distribution is nearly Gaussian, and hence the ratio has a plateau in the low $k_\perp$ region at $R_N \simeq (Q_S^p)^2/(Q_S^A)^2 \simeq 1/3$. For high $k_\perp$, $R_N \to 1$ as a consequence of the perturbative tail. Going lower in $x$ via the rcBK equation, the peak position moves toward the high-$k_\perp$ region and the plateau shrinks---already for $x \sim 0.001$ the value of $1/3$ is reached only for very small $k_\perp$. For $k_\perp$ as large as 2.9 GeV, $R_N$ never deviates significantly from unity. Therefore, even in this extreme scenario $A_N$ does not have nuclear dependence. This is also illustrated by the thin curves in Fig.~\ref{fig:phenix} where we computed $A_N$ including only the derivative term in the numerator. 
 Note that we have only included the fragmentation contribution, but it is clear that adding the contribution from the twist-three quark-gluon distributions \cite{Hatta:2016wjz} will not help resolve this issue.

\section{Conclusions}

We have made a numerical computation of SSA in $p^\uparrow p$ and $p^\uparrow A$ in the forward region including the gluon saturation effect of the nucleus. Using the current state-of-the-art twist-3 fragmentation functions and the dipole gluon distribution, we compared our results to the preliminary STAR and PHENIX data. While the saturation based description seems to describe well the overall magnitude of the STAR data, it fails to explain the scaling $A_N\sim A^{-1/3}$ observed by  the PHENIX collaboration.

According to the result of \cite{Hatta:2016wjz,Hatta:2016khv}, a strong nuclear suppression of $A_N$ is possible only if the $\sim A^{-1/3}$ terms  dominate over the  $\sim A^{0}$ terms, and if one looks at $P_{h\perp}$ less than $Q_S$.   The recent fit of the $pp$ data  \cite{Gamberg:2017gle} suggests that the first condition does not hold, 
and the second condition is also violated by the high value of $P_{h\perp}$ measured by the PHENIX collaboration. This makes the PHENIX result all the more striking. It is also puzzling that there seems to be a sudden change in the behaviors of $A_N$ between $x_F=0.2$ (the lowest value measured by the STAR collaboration) and $x_F=0.12$ (highest value measured by the PHENIX collaboration). 

 This may call for alternative mechanisms of SSA around $x_F\sim 0.1$ whose nuclear dependence come from a different source. Indeed the region $0.1\lesssim x_F\lesssim 0.2$ might be special---it is roughly the `threshold' region where $A_N$ starts to grow. Thus the value of $A_N$ itself is very small in this region and a small effect can cause a large numerical impact. Perhaps one should include not only the $q+g$ channel (as we did in this paper), but also the $g+g$ channel. Since there is no `gluon transversity' distribution, the fragmentation contribution is absent in this channel, but the collinear three-gluon, or `odderon' contribution comes into play \cite{Koike:2011mb,Koike:2011ns,Kovchegov:2012ga,Hatta:2013wsa,Zhou:2015ima}. 
 A precise evaluation of SSA including these effects appears to be a challenging task.

\section*{Acknowledgments}
We thank Carl Gagliardi, and  Jeongsu Bok, Jin Huang for sharing with us the preliminary STAR and PHENIX data, respectively. We thank Kevin Dusling for tables of the rcBK solution and Alexey Prokudin for transversity and fragmentation functions as well as the clarifications of the uncertainty analysis. We also thank Elke Aschenauer for comment on uncertainty in transversity and fragmentation functions.
Y.~H. is supported by the U.S. Department of Energy, Office of Science, Office of Nuclear Physics, under Contracts No. de-sc0012704. S.~B. is supported by a JSPS postdoctoral fellowship for foreign researchers under Grant No.~17F17323. S.~B. also acknowledges HRZZ Grant No. 8799 for computational resources.

\bibliographystyle{h-physrev}
\bibliography{referencescgc}

\end{document}